\documentclass[prd,twocolumn,floatfix,preprintnumbers,letterpaper]{revtex4}

\usepackage{graphicx}
\input{epsf}
\usepackage{epsf}
\usepackage{graphicx,epsfig}
\usepackage{bm}
\usepackage{latexsym,amssymb,amsmath,float}

\newcommand {\ga} {\ {\raise-.5ex\hbox{$\buildrel>\over\sim$}}\ }
\newcommand {\la} {\ {\raise-.5ex\hbox{$\buildrel<\over\sim$}}\ } 
\def\be{\begin{equation}}
\def\ee{\end{equation}}

\begin{document}

\title{Hilltop Quintessence}
\author{Sourish Dutta and Robert J. Scherrer}
\affiliation{Department of Physics and Astronomy, Vanderbilt University,
Nashville, TN  ~~37235}

\begin{abstract}
We examine hilltop quintessence models, in which
the scalar field is rolling near a local maximum in the potential,
and $w \approx -1$.  We first derive
a general equation for the evolution of $\phi$ in the limit
where $w \approx -1$.  We solve this equation for the case
of hilltop quintessence
to derive $w$ as a function of the scale factor; these solutions depend on
the curvature of the potential near its maximum.
Our general result is in excellent agreement ($\delta w \alt 0.5\%$)
with all of the particular cases examined.  It works particularly well
($\delta w \alt 0.1\%$) for the pseudo-Nambu-Goldstone Boson potential.
Our expression for $w(a)$ reduces to the previously-derived
slow-roll result of Sen and Scherrer in the limit where the curvature
goes to zero.  Except for this limiting case, $w(a)$ is poorly
fit by linear evolution in $a$.

\end{abstract}

\maketitle

\section{Introduction}

Observational evidence \cite{Knop,Riess1}
suggests that approximately 70\% of the energy density in the
universe is in the form of an exotic, negative-pressure component,
dubbed dark energy.  (For a recent
review, see, e.g., Ref. \cite{Copeland}).
The observational bounds on the properties
of the dark energy have continued to tighten.  Taking
$w$ to be the ratio of
pressure to density for the dark energy:
\begin{equation}
\label{w}
w = p_{DE}/\rho_{DE},
\end{equation}
recent observational constraints are typically
$-1.1 \la w \la -0.9$ when $w$ is assumed constant
(see, e.g., \cite{Wood-Vasey,Davis}
and references therein).

One possibility is that the dark energy is, in fact,
merely a cosmological constant, with $w$ exactly equal
to $-1$.  This is the standard
$\Lambda$CDM model, in which
the universe today is roughly 30\% matter, and
70\% vacuum energy, with the latter having a constant density.
In this case, we expect future observations
to continue to converge on this value of $w$.
A second possibility is
that $w$ is close to $-1$, but not exactly equal to it.
In this case, an alternative explanation for the
dark energy is required.

One possible model, dubbed quintessence, assumes that
the dark energy arises from a minimally coupled
scalar field; these
models have been extensively explored.
(See, for instance, Refs. \cite{ratra,turner,caldwelletal,liddle,zlatev}
for some of the earliest discussions).  The
equation of motion for the scalar field simplifies
considerably in the limit where the background expansion
is well-described by $\Lambda$CDM, which is the case whenever
$w$ for the quintessence field is always close to $-1$.
This fact was exploited in Ref. \cite{SS}, which considered quintessence
models in a potential satisfying the slow-roll conditions,
i.e.,
\begin{equation}
\label{slow1}
\left(\frac{1}{V} \frac{dV}{d\phi}\right)^2 \ll 1,
\end{equation}
and
\begin{equation}
\label{slow2}
\frac{1}{V}\frac{d^2 V}{d\phi^2} \ll 1.
\end{equation}
It was shown in this paper that the evolution of $w(a)$ in these models,
which we will call ``slow-roll quintessence,"
approaches a single functional form, given by the present-day
values of $w$ and $\Omega_\phi$.  (Note that the term
``slow-roll quintessence" refers to the fact that the potential
satisfies the slow-roll conditions; the well-known slow-roll
approximation for inflation cannot be applied \cite{SS}.
For alternative approaches to
this problem, see Refs. \cite{Crit,Neupane,Cahn}).

However, the slow-roll conditions given by Eqs. (\ref{slow1})
and (\ref{slow2}), while sufficient to give $w$ near $-1$ today,
are not necessary.  In this paper, we consider a second possibility:
a scalar field rolling down near a local maximum in the potential.
We assume that we are sufficiently close to the maximum that
Eq. (\ref{slow1}) still applies, but we relax our assumption
in Eq. (\ref{slow2}).  In analogy with recent discussions
of similar models in inflation \cite{hill1,hill2,hill3}, we call these models
``hilltop quintessence."  A particularly important model that
is well-described by this methodology is the pseudo-Nambu-Goldstone Boson
(PNGB) model \cite{Frieman}.

In the next section, we derive the version of the scalar field
equation of motion that applies in the limit where the background
expansion is well-approximated by a $\Lambda$CDM model.  In
Sec. III, we expand
the potential for hilltop quintessence
as a quadratic function near its maximum and
solve this equation to derive the evolution of the scalar
field and thus, the evolution of its equation of state, for
generic hilltop quintessence models.
We find that all of these models can be characterized by a single
set of functional forms for $w(a)$ that depends on the present-day values
of $\Omega_\phi$ and $w$ for the quintessence, along with a parameter ($K$)
that depends on the curvature of the scalar field potential
at its maximum.  This result is given in Eq. (\ref{finalfinal}).
Our results are discussed in Sec. IV.

\section{Scalar field evolution in the $\Lambda$CDM limit}

Consider a minimally-coupled
scalar field, $\phi$, in a potential $V(\phi)$.
The density and pressure of the
scalar field are given by
\begin{equation}
\label{density}
\rho = \frac{\dot \phi^2}{2} + V(\phi),
\end{equation}
and
\begin{equation}
\label{pressure}
p = \frac{\dot \phi^2}{2} - V(\phi),
\end{equation}
respectively, and the equation of state parameter, $w$,
is given by equation (\ref{w}).
The equation of motion for this field in an expanding background is
\begin{equation}
\label{motionq}
\ddot{\phi}+ 3H\dot{\phi} + \frac{dV}{d\phi} =0,
\end{equation}
where the Hubble parameter $H$ is given by
\begin{equation}
\label{H}
H = \left(\frac{\dot{a}}{a}\right) = \sqrt{\rho_T/3},
\end{equation}
and we have assumed a flat universe.
Here $a$ is the scale factor, $\rho_T$ is the total density, and we work in units
for which $8 \pi G = 1$.  The evolution of the scale factor is further
described by
\begin{equation}
\label{ddota}
\frac{\ddot a}{a} = -\frac{1}{6}(\rho_T + 3 p_T),
\end{equation}
where $p_T$ is the total pressure.

We first eliminate the $\dot \phi$ term in Eq. (\ref{motionq})
by making the change of variables
\begin{equation}
\phi(t) =  u(t)/a(t)^{3/2},
\end{equation}
which gives
\begin{equation}
\ddot u - \frac{3}{2}\left[\frac{\ddot a}{a} + \frac{1}{2} \left(
\frac{\dot a}{a}\right)^2
\right]u + a^{3/2} V^\prime(u/a^{3/2})=0.
\end{equation}

Applying the expressions for $\dot a/a$ and $\ddot a/a$
from Eqs. (\ref{H}) and (\ref{ddota}) gives
\begin{equation}
\label{exact}
\ddot u + \frac{3}{4} p_T u + a^{3/2}V^\prime(u/a^{3/2}).
\end{equation}
This equation is as yet exact.

Eq. (\ref{exact}) takes a particularly simple form
for two special cases:  a matter-dominated universe (for which
$p_T = 0$) and a $\Lambda$CDM universe (for which $p_T$ is constant).
In the former case, we obtain
\begin{equation}
\ddot u + tV^\prime(u/t) = 0,
\end{equation}
and it is easy to use this equation to derive, in a straightforward
way, the results of Refs. \cite{ratra} and \cite{liddle}.  However,
the subject of this paper is the second case.

We assume a universe
containing matter and quintessence, but such that the quintessence
always has $w$ near $-1$.  In this limit, $p_T$ is well-approximated
by a constant:
$p_T \approx -\rho_{\phi 0}$, where $\rho_{\phi 0}$ is the nearly constant
density contributed by the quintessence in this limit.  
Then Eq. (\ref{exact}) becomes
\begin{equation}
\label{lambda_u}
\ddot u - \frac{3}{4}\rho_{\phi 0}u + a^{3/2}V^\prime(u/a^{3/2}) = 0.
\end{equation}
We expect Eq. (\ref{lambda_u}) to provide a good approximation to
the evolution of any scalar field in the limit where
the Hubble parameter in Eq. (\ref{motionq}) is well-approximated
by $\Lambda$CDM, i.e., any model in which the
kinetic term in Eqs. (\ref{density}) and (\ref{pressure}) is
dominated by the potential term, so that $w$ for the scalar field
never evolves very far from $-1$.  Eq. (\ref{lambda_u}) applies, for example,
to the limiting model examined in Ref. \cite{SS}.  We now apply
it to a different class of models.

\section{Hilltop scalar field evolution}

We are interested in models in which the scalar field evolves
near a local maximum in the potential.  The most important
model of this type is the PNGB model \cite{Frieman}, for which
the potential is given by
\begin{equation}
\label{PNGB}
V(\phi) = M^4 [\cos(\phi/f)+1],
\end{equation}
where $M$ and $f$ are constants.  (For
recent discussions of the PNGB
model in the context of dark energy, see, e.g., Refs. \cite{Dutta,Albrecht,LinderPNGB}
and references therein).
Other, less-well-motivated models with a local maximum in the potential
include the Gaussian potential,
\begin{equation}
\label{Gaussian}
V(\phi) = M^4 e^{-\phi^2/\sigma^2},
\end{equation}
and the quadratic potential
\begin{equation}
\label{quadratic}
V(\phi) = V_0 - V_2\phi^2.
\end{equation}
Our purpose in this paper, however, is not to examine an exhaustive
list of such models, but to show that they all converge
to a common evolution under certain conditions.

We note that any model with a local maximum in the potential
at $\phi = \phi_*$
can be expanded, near this maximum, in the form
\begin{equation}
\label{expand}
V(\phi) = V(\phi_*) + (1/2)V^{\prime \prime}(\phi_*) \phi^2 + O(\phi^3) +...
\end{equation}
This expansion will be a good approximation for the PNGB model when
$\phi \ll f$ and for the Gaussian model
as long as $\phi \ll \sigma$.  It is 
exact for the quadratic potential.  The evolution of $\phi$
for this expansion (or, equivalently, for the quadratic potential
of Eq. \ref{quadratic}) was previously examined
in Refs. \cite{Alam,Linde1,Linde2,Linde3}.  The motivation in
Ref. \cite{Alam} was similar to our own investigation, while Refs. \cite{Linde1,
Linde2,Linde3} were concerned with the future fate of the universe
in such a model.  The exact solution to the scalar field equation
of motion for this potential is given in Ref. \cite{Alam} for the
matter-dominated case, while Refs. \cite{Linde1,Linde2,Linde3} give
the evolution for the scalar-field-dominated case (appropriate for
the far future of the universe).  Here we solve for the general case
in which both matter and the scalar field contribute to the evolution,
since this is the relevant case at low redshift.

Substituting
the expansion of Eq. (\ref{expand}) into Eq. (\ref{lambda_u}) (and
taking $\rho_{\phi 0} = V(\phi_*)$) gives
\begin{equation}
\label{eqfinal}
\ddot u + [V^{\prime \prime}(\phi_*) - (3/4)V(\phi_*)]u = 0.
\end{equation}
If we define the constant $k$ to be given by
\begin{equation}
k \equiv  \sqrt{(3/4)V(\phi_*) - V^{\prime \prime}(\phi_*)},
\end{equation}
then the general solution to Eq. (\ref{eqfinal})
is
\begin{equation}
u = A \sinh(kt) + B\cosh(kt).
\end{equation}
Now we assume that the scale factor is well-approximated
by its value in the $\Lambda$CDM model, which is again
a consequence of the assumption that $w$ is always close to $-1$:
\begin{equation}
\label{LCDM a}
a(t)=\left[ \frac{1-\Omega_{\phi 0}}{\Omega_{\phi 0}}
\right]^{1/3}\sinh^{2/3}(t/t_\Lambda),
\end{equation}
where $\Omega_{\phi 0}$ is the present-day value of $\Omega_\phi$,
and $a=1$ at present.  The
time $t_\Lambda$ is defined to be
\begin{equation}
t_\Lambda = 2/\sqrt{3 \rho_{\phi 0}} = 2/\sqrt{3 V(\phi_*)}.
\end{equation}
With this expression for $a(t)$, the general solution for
$\phi(t)$ is given by
\be
\label{hilltop_phi}
\phi(t)=\left[\frac{\Omega_{\phi 0}}{1-\Omega_{\phi 0}} \right]^{1/2}
\frac{A\sinh(kt)+B\cosh(kt)}{\sinh(t/t_\Lambda)}
\ee
We require that at $t=0$, $\phi$ is equal to a fixed initial value, $\phi_i$.
This forces $B=0$ and gives us the value of $A$, so that
\be
\label{phievol}
\phi = \frac{\phi_i}{kt_\Lambda} \frac{\sinh(kt)}{\sinh(t/t_\Lambda)}.
\ee
In the limit where $t \ll t_\Lambda$, our solution reduces to the
matter-dominated solution in Ref. \cite{Alam}, while for $t \gg t_\Lambda$,
we regain the scalar-field-dominated solution of Refs. \cite{Linde1,
Linde2,Linde3}.

The equation of state parameter for quintessence is given by
\begin{equation}
\label{state}
1+w = \frac{\dot \phi ^2}{\rho_\phi}.
\end{equation}
Taking $\rho_\phi \approx \rho_{\phi 0} \approx V(\phi_*)$, Eqs. (\ref{phievol})
and (\ref{state}) yield
\begin{widetext}
\begin{equation}
1+w = \frac{3}{4} \frac{\phi_i^2}{k^2} \left[\frac{k \cosh(kt)\sinh(t/t_\Lambda) - (1/t_\Lambda)
\sinh(kt)\cosh(t/t_\Lambda)}{\sinh^2(t/t_\Lambda)}\right]^2.
\end{equation}
\end{widetext}
We can normalize this expression to the present-day value of $w$, which we denote $w_0$,
and we can use Eq. (\ref{LCDM a}) to express $w$ as a function of the scale factor
(or the redshift) rather than $t$.  We obtain:
\begin{widetext}
\be
\label{final EOS}
1+w(a)=(1+w_0)a^{-3}
\frac{{\left[ {\sqrt {\Omega _{\phi 0}} kt_\Lambda\cosh
\left[ {k t\left( a \right)} \right] - \sqrt {(1-\Omega_{\phi 0})a^{-3}
+\Omega_{\phi 0}}
\sinh \left[ {k t\left( a \right)} \right]} \right]^2 }}
{{\left[ {\sqrt {\Omega_{\phi 0}}  kt_\Lambda\cosh ( {kt_0})
 - \sinh ( {k t_0})} \right]^2 }},
\ee
\end{widetext}
where $t(a)$ and $t_0$ can be derived from Eq. (\ref{LCDM a}):
\begin{equation}
\label{ta}
t(a) = t_\Lambda \sinh^{-1}\sqrt{\left(\frac{\Omega_{\phi 0} a^3}
{1-\Omega_{\phi 0}}\right)}
\end{equation}
and
\begin{equation}
\label{t0}
t_0 = t_\Lambda \tanh^{-1} \left(\sqrt{\Omega_{\phi 0}}\right).
\end{equation}
We now define the constant $K \equiv kt_\Lambda$.  In
terms of the quintessence potential, $K$ is just
\begin{equation}
\label{Kdef}
K = \sqrt{1 - (4/3)V^{\prime \prime}(\phi_*)/V(\phi_*)}.
\end{equation}
Thus, $K$ depends only on the value of the potential and its
second derivative at its maximum.
Note that $V^{\prime \prime}(\phi_*) < 0$, so $K > 1$.
Then Eq. (\ref{final EOS}) can be written as
\begin{widetext}
\begin{equation}
\label{finalfinal}
1 + w(a) = (1+w_0)a^{3(K-1)}\frac{[(F(a)+1)^K(K-F(a))
+(F(a)-1)^K(K+F(a))]^2}
{[(\Omega_{\phi0}^{-1/2}+1)^K(K-\Omega_{\phi0}^{-1/2})
+(\Omega_{\phi0}^{-1/2}-1)^K (K+\Omega_{\phi0}^{-1/2})]^{2}},
\end{equation}
\end{widetext}
where $F(a)$ is given by
\be
F(a) = \sqrt{1+(\Omega_{\phi 0}^{-1}-1)a^{-3}}.
\ee
(Note that $F(a) = 1/\sqrt{\Omega_\phi(a)}$, where $\Omega_\phi(a)$
is the value of $\Omega_\phi$ as a function of redshift, so that
$F(a=1) = \Omega_{\phi0}^{-1/2}$.)

Eq. (\ref{finalfinal})
is our main result.  It shows that, in the limit where $w$
is close to $-1$ (i.e., the scalar field potential energy
dominates the kinetic energy), all of the hilltop quintessence
models with a given value of $w_0$ form a single set of models
parametrized by the value of $K$, which depends only
on $V^{\prime \prime}(\phi_*)/V(\phi_*)$.  For the case
of slow-roll quintessence models, a similar analysis yields
only a single form for $w(a)$, once $w_0$ and $\Omega_{\phi 0}$
are fixed \cite{SS}.  Here we have more complex behavior, since $w(a)$
also varies with $K$.  This behavior is illustrated in
Fig. 1, where we fix $w_0 = -0.9$ and $\Omega_{\phi 0} = 0.7$
and then plot $w(a)$ from Eq. (\ref{finalfinal}) as a function
of $a$.
\begin{figure}[t]
	\epsfig{file=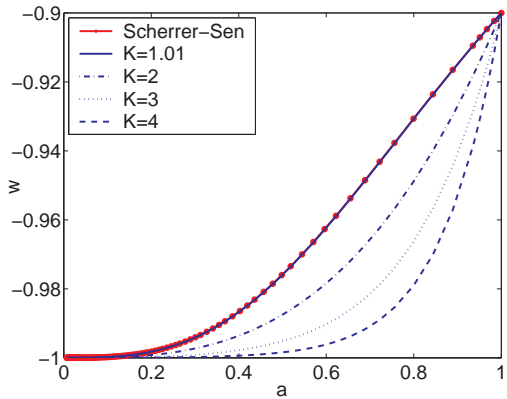,height=55mm}
	\caption
	{
The evolution of $w(a)$ given by Eq. \eqref{finalfinal} for hilltop
quintessence models with $w_0 = -0.9$ and $\Omega_{\phi 0} = 0.7$,
for the indicated values
of $K$, as defined in Eq. \eqref{Kdef}.  Red curve (filled circles)
gives the approximation from Ref. \cite{SS} for slow-roll
quintessence.}
\end{figure}
When $V^{\prime \prime}(\phi_0)/V(\phi_0) \gg 1$, we see
that the field evolution diverges significantly from slow-roll
quintessence.  In particular, $w$ increases more slowly at high redshift, but
much more rapidly at low redshift.  This is easily understood from the nature
of these potentials:  $dV/d\phi$ increases as $\phi$ rolls down
the potential.  In the limit where $V^{\prime\prime}(\phi_0)/V(\phi_0)
\rightarrow 0$, corresponding to $K \rightarrow 1$, both
slow-roll conditions (Eqs. \ref{slow1} and \ref{slow2})
apply, and we expect the evolution to converge
to the form derived in Ref. \cite{SS}, namely
%\begin{widetext}
\begin{equation}
\label{wpred}
1 + w = (1+ w_0)\frac{\Biggl[ F(a)
- \left[F(a)^2-1\right] \tanh^{-1} \left[F(a)^{-1}\right]
\Biggr]^2}
{ \left[\Omega_{\phi 0}^{-1/2}
- \left(\Omega_{\phi 0}^{-1} - 1 \right)
\tanh^{-1}\sqrt{\Omega_{\phi 0}}\right]^{2}}.
\end{equation}
%\end{widetext}
It is straightforward to show that Eq. (\ref{finalfinal})
reduces to Eq. (\ref{wpred}) in the limit where $K \rightarrow 1$.
We also show this effect in Fig. 1:  the curve corresponding to $K = 1.01$
is indistinguishable from $w(a)$ for slow-roll
quintessence.

Eq. (\ref{finalfinal}) simplifies considerably for the case where
$K$ is a small integer.  We have, for example,
\begin{equation}
K=2:~~~~~~1+w = (1+w_0) a^3,
\end{equation}
\be
K=3:~~~~~~1+w = (1+w_0)[(1-\Omega_{\phi0})a^3 + \Omega_{\phi0}a^6],
\ee
\begin{widetext}
\be
K=4~~~~~~1+w = \frac{1+w_0}{(5+\Omega_{\phi0})^2}\left[
25(1-\Omega_{\phi0})^2 a^3 + 60\Omega_{\phi0}(1-\Omega_{\phi 0})a^6
+ 36 \Omega_{\phi 0}^2 a^9 \right].
\ee
\end{widetext}
Of course, it would require rather bizarre fine-tuning of the
potential for $K$
to be exactly equal to one of these values, but these special
cases provide some qualitative insight into the behavior of $w(a)$
as a function of $K$.

Now we evaluate the accuracy of Eq. (\ref{finalfinal}) when applied
to various models of interest.  Consider the three hilltop potentials
outlined above.  We have, for the quadratic potential,
$K = \sqrt{1 + (8/3)(V_2/V_0)}$, for the PNGB potential,
$K = \sqrt{1+(2/3)(1/f^2)}$, and for the Gaussian potential,
$K = \sqrt{1+(8/3)(1/\sigma^2)}$.  In Figs. 2-4, we have plotted our
expression for $w(a)$ from Eq. (\ref{finalfinal}) against the exact numerical evolution for these
three potentials, fixing $\Omega_{\phi0} = 0.7$.  (Once the potential
and $\Omega_{\phi 0}$ are fixed, the value of $\phi_i$ is chosen to produce the desired value
of $w_0$).
\begin{figure}[t]
	\epsfig{file=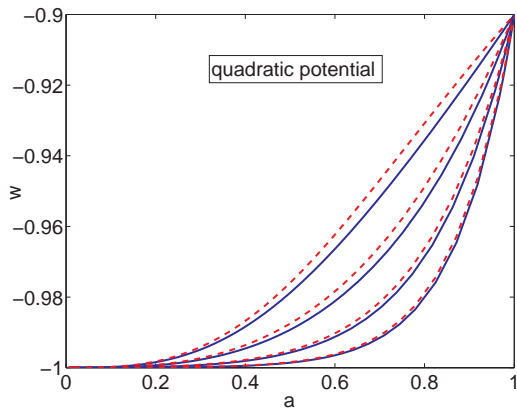,height=55mm}
	\caption
	{Comparison between our approximation for $w(a)$ for
	hilltop quintessence (Eq. \ref{finalfinal}) with
	$w_0 = -0.9$ and $\Omega_{\phi 0} = 0.7$
	and the exact (numerically-integrated) evolution for $w(a)$ for
	the quadratic potential $V = V_0 - V_2 \phi^2$.  Red (dashed)
	curves give our approximation, and solid (blue) curves
	give exact evolution, for (left to right),
	$K = 1.01, 2, 3, 4$, where $K$ is defined
	by Eq. \eqref{Kdef}.
	}
\end{figure}
\begin{figure}[t]
	\epsfig{file=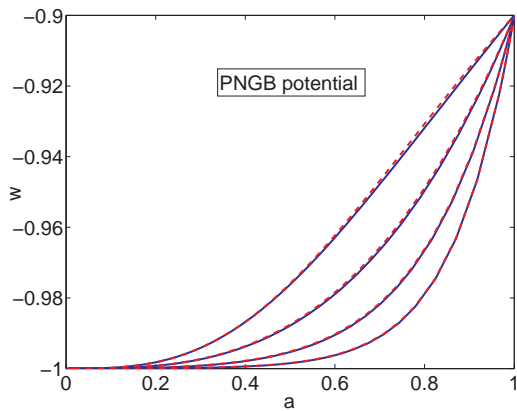,height=55mm}
	\caption
	{As Fig. 2, for the PNGB potential, $V = M^4[\cos(\phi/f)+1]$.}
\end{figure}
\begin{figure}[t]
	\epsfig{file=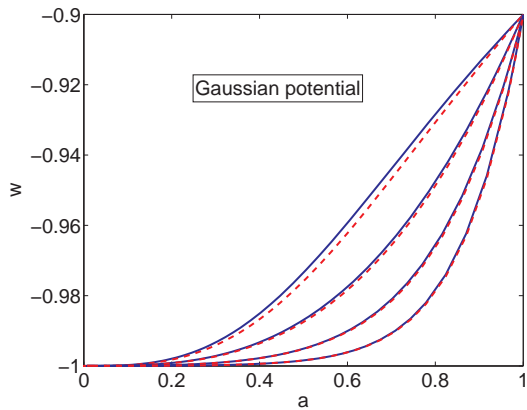,height=55mm}
	\caption
	{As Fig. 2, for the Gaussian potential, $V = M^4 e^{-\phi^2/\sigma^2}$.}
\end{figure}

The agreement, in all three cases, between Eq. (\ref{finalfinal}) and
the exact numerical evolution is excellent, with errors $\delta w \alt 0.5\%$
for the quadratic potential, $\delta w \alt 0.1\%$ for the
PNGB potential, and $\delta w \alt 0.3\%$ for the
Gaussian potential.
Rather surprisingly, the errors for the quadratic potential (which is
the basis of our approximation) are actually larger than for the other
two potentials.  However, this is due to the fact that we have made
several approximations in our calculation; these tend to cancel for
the PNGB and Gaussian potentials.

The agreement between Eq. (\ref{finalfinal}) and the exact evolution
is particularly striking for the case of the PNGB potential, which also
happens to be the most interesting and
well-motivated of the hilltop potentials.  We have
therefore examined this potential in more detail.  In Fig. 5,
we extend our results to larger values of $w_0$.  It is interesting to
see that the agreement remains excellent for $w_0$ as large as $-0.7$.
Thus, Eq. (\ref{finalfinal}) represents a nearly exact solution for
$w(a)$ for the PNGB model within a wide range for $w_0$.
\begin{figure}[t]
	\epsfig{file=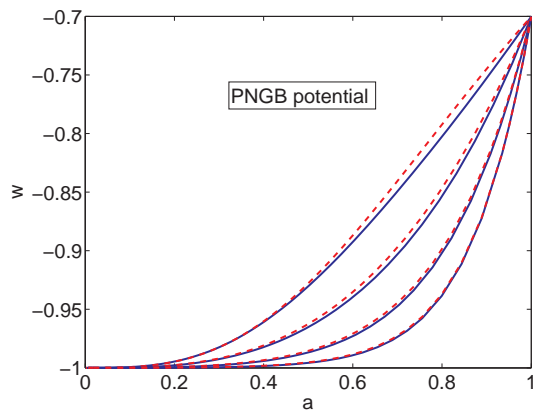,height=55mm}
	\caption
	{As Fig. 3, for the PNGB potential, extended to larger values of $w_0$.}
\end{figure}
\begin{figure}
	\epsfig{file=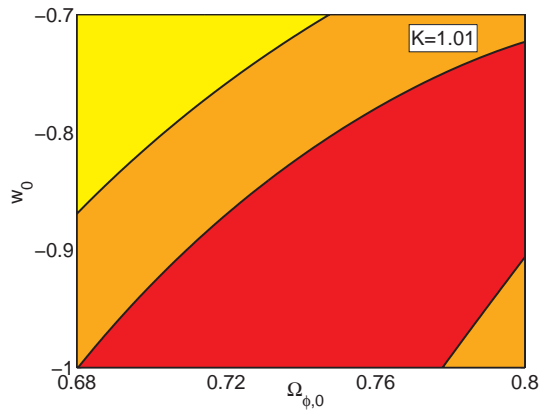,height=55mm}
	\caption
	{\label{likelihood}
Likelihood plot from SNIa data for the parameters $w_0$ and $\Omega_{\phi0}$,
for hilltop quintessence models with generic behavior
described by Eq. \eqref{finalfinal}, with $K=1.01$,
where $K$
is the function of the curvature of the potential
at its maximum given in Eq. \eqref{Kdef}.
The yellow (light) region
is excluded at the 2$\sigma$ level, and the orange (darker) region
is excluded at the 1$\sigma$ level.  Red (darkest) region is
not excluded at either confidence level.}
\end{figure}
\begin{figure}
	\epsfig{file=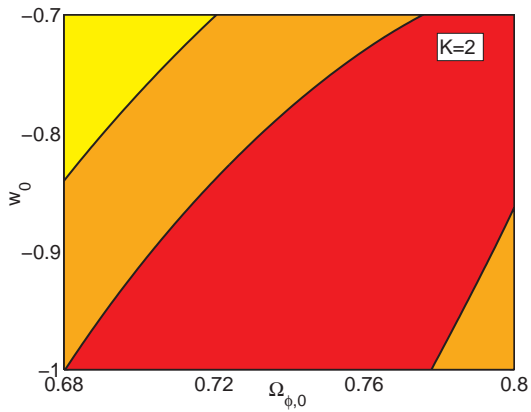,height=55mm}
	\caption
	{\label{likelihood2}
As Fig. 6, for $K=2$.}
\end{figure}\begin{figure}
	\epsfig{file=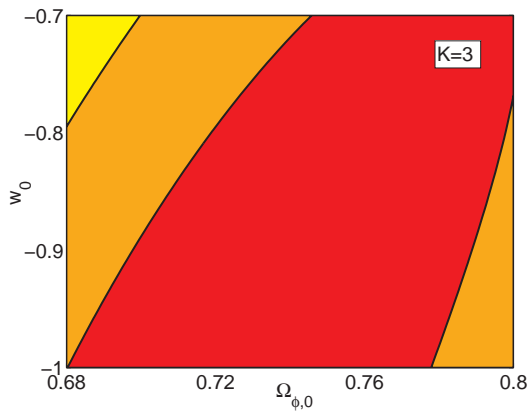,height=55mm}
	\caption
	{\label{likelihood3}
As Fig. 6, for $K=3$.}
\end{figure}\begin{figure}
	\epsfig{file=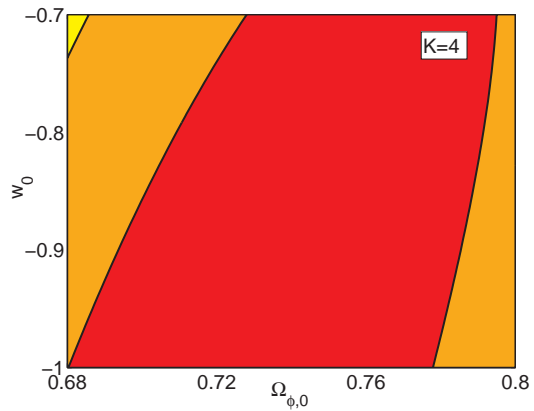,height=55mm}
	\caption
	{\label{likelihood4}
As Fig. 6, for $K=4$.}
\end{figure}

Finally, in Figs. 6-9,
we use Eq. \eqref{finalfinal}
to construct a $\chi^2$ likelihood plot for $w_0$
and $\Omega_{\phi0}$ with $K = 1.01, 2, 3, 4$,
using the recent Type Ia Supernovae
standard candle data (ESSENCE+SNLS+HST from \cite{Davis}).
While none of these models is ruled out by current supernova data,
it is interesting to note that the hilltop quintessence models
($K = 2,3,4$) produce a larger allowed region than the slow-roll
quintessence model ($K = 1.01$), and the allowed region increases
with increasing $K$.

\section{Discussion}

Our results indicate that hilltop quintessence models with $w$ near $-1$
all produce a similar evolution for $w(a)$, given by Eq. (\ref{finalfinal}).
The importance of this result lies in the fact that a very
general set of models can
be mapped onto a fairly constrained set of behaviors for $w(a)$.
Note that in general, the evolution given by Eq. (\ref{finalfinal})
and shown in Fig. 1 is {\it not}
well-described by the popular linear parametrization,
$w(a) = w_0 + w_a(1-a)$ \cite{CPL}.  The one exception is
the limiting case $K \rightarrow 1$, where we
regain the results of Ref. \cite{SS}: in this limit
the evolution is roughly linear for $a > 0.5$ ($z < 1$),
with $w_a \approx -1.5(1+w_0)$.

This investigation, along with Ref. \cite{SS}, can also be thought of
as a kind of
Taylor expansion of the potential about the initial
value of $\phi$.  The results of Ref. \cite{SS}
apply when the linear term dominates,
while the results presented here assume a quadratic expansion.
The results of Ref. \cite{SS} are then a special case (albeit a very
important special case) of the results presented here.

Of course, our results do not apply to all thawing quintessence models
with $w$ near $-1$,
but only those satisfying Eq. (\ref{slow1}).  A different approach
was taken recently by Cahn, et al. \cite{Cahn}.  They looked at the
evolution of the scalar field at early times, when
$\rho_T \gg \rho_\phi$, while assuming nothing
about the detailed nature of the potential.  This allows the derivation
of a generic result for the evolution of the scalar field
before it dominates the expansion (see also Ref. \cite{Watson} for the special case of tracker fields).
Our results for the particular class of potentials
considered here agree with those given in Ref. \cite{Cahn} (as they should) in the limit
where $\Omega_\phi \ll 1$.

\acknowledgments

R.J.S. was supported in part by the Department of Energy (DE-FG05-85ER40226).
We thank A.A. Sen for helpful discussions.

\end{document}